\documentclass[letterpaper, reqno]{amsart}
\usepackage{epsfig}
\usepackage{oldlfont}

\newcommand{\ket}[1]{\mid #1 \rangle}
\newcommand{\braket}[2]{\langle #1 \mid #2 \rangle}

\newcommand{\pic}[5]{\raisebox{#3pt}
{\hspace{#4pt} \epsfig{file=#1.ps,height=#2pt,silent=} 
\hspace{#5pt}}}

\begin{document}

\title{New Operators for Spin Net Gravity: Definitions and Consequences}

\author{Seth A. Major}

\date{January 2001}
\address{Department of Physics, Hamilton College,
Clinton NY 13323 USA}
\email{smajor@hamilton.edu}

\begin{abstract}
Two operators for quantum gravity, angle and quasilocal energy, are
briefly reviewed.  The requirements to model semi-classical angles are
discussed.  To model semi-classical angles it is shown that the internal
spins of the vertex must be very large, $\sim 10^{20}$.

\end{abstract}

\maketitle

\section{Introduction}

Spin net gravity is based on the quantization of gravity using
connection variables.  (For a review, see Ref.  \cite{carlorev}.) 
Searching for the eigenspace for geometric observables, Rovelli and
Smolin introduced new states \cite{rsops, rsspinet} which echoed an
older, combinatorial definition of spacetime advocated by Penrose
\cite{penrose}.  Using Penrose's original name the new states were
named ``spin networks''.  These were later shown to be a basis
\cite{baez} for the kinematic or 3-geometry states of quantum gravity. 
But despite this success, the dynamics remains controversial.  ``Spin
net gravity'' makes use of the kinematic states and includes the
assumption that the spin network states are robust enough to support
the full dynamics of quantum gravity.

Spin networks are discrete, graph-like structures.  A spin network,
${\cal N}$, is defined as the triple $(\mathsf{G}; {\bf i, n})$ of an
oriented graph, $\mathsf{G}$, labels on the vertices (or
``intertwiners''), ${\bf i}$, and integer edge labels, ${\bf n}$.  The
edge labels index the irreducible representation carried on the edge. 
The corresponding spin net state $\ket{s}$ is defined in the
connection representation as
\[
\braket{A}{s} 
\equiv \braket{A}{ {\mathsf G} \, {\bf  i \,  n} } 
:= \prod_{v \in {\bf v}( {\mathsf G} ) }
{\bf i}_v  \circ \otimes_{e \in e({\mathsf G})} U_{e}^{(n_{e})}[A],
\]
where the holonomy $U_{e}^{(n_{e})}$ along edge $e$ is in the
$n_{e}/2$ irreducible representation of $SU(2)$.  When the
intertwiners ``connect'' all the incident edges.  When the
intertwiners are invariant tensors on the group, the states are
gauge invariant.  One may represent intertwiners of higher valence
vertices as a decomposition into trivalent vertices and labels on the
``internal'' edges which connect these trivalent vertices.

Spin networks are the eigenspace of quantum operators which measure
geometric quantities such as length, area, volume, and angle.  For
instance, when an edge $e$ of a spin network passes through a surface,
it contributes area $\ell^{2} \sqrt{n_{e} (n_{e}+2)}$ to the area of
that surface \cite{rsops}.  Hence, edges are called ``flux lines of
area''.  Vertices support volume and angle and may be called ``atoms
of geometry''.  The new quantum geometry is completely different than
our everyday, continuous geometry.  In fact, this theory predicts that
space is fundamentally discrete.

In the next section an angle operator is introduced and its
semi-classical limit is discussed.  A quasilocal energy operator is
briefly reviewed in Section \ref{quasi}.  The report is summarized in
Section \ref{sum}.

\section{Angle Operator}

The idea that one could measure angle from the discrete structure of
spin networks first arose in the context of Penrose's work
\cite{penrose}.  This work culminated in the Spin Geometry Theorem,
which stated that angles in 3-dimensional space could be approximated
to arbitrary accuracy if the spin network was sufficiently correlated
and if the spins were sufficiently large.  This result is an extension
of the observation that, for an EPR pair, we know everything about the
relative orientation of the particles' spins but nothing about the
absolute orientation of the two spins in space.  The spin geometry
construction carries over into the context of quantum geometry: One
can measure the angle between two (internal) edges of a spin network
state.

\begin{figure} 
    \begin{center}
	\pic{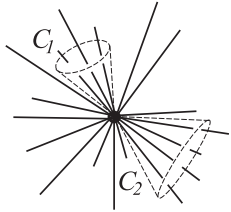}{80}{-10}{-1}{2}
    \end{center}
		 \caption{ \label{cones} A vertex with two bundles of
		 edges defined by the cones $C_{1}$ and $C_{2}$.  The
		 bases of the two cones are surfaces denoted $S_{1}$
		 and $S_{2}$, respectively.  The core of the
		 intertwiner for the vertex is chosen so that all
		 edges passing through $S_{1}$ ($S_{2}$) connect to a
		 single internal edge $n_{1}$ ($n_{2}$).}
\end{figure}

The angle operator is a quantization of the classical expression for 
an angle $\theta_{v}$ measured at a point $v$
\begin{equation}
    \theta^{(12)}_{v} = \arccos \frac{E_{S_{1}}^{i} E_{S_{2}}^{i}}
{\sqrt{E_{S_{1}}^{i} E_{S_{1}}^{i}} \sqrt{E_{S_{2}}^{j} E_{S_{2}}^{j}}}
= \arccos \frac{ q \, q^{ab} n_{a} m_{b} }
	{\sqrt{ q q^{ab} n_{a} n_{b} } \, 
	\sqrt{q q^{cd} m_{c} m_{d}}}
\end{equation}
with surface normals $n$ and $m$ for surfaces $S_{1}$ and $S_{2}$. The
regularization of the classical expression is straightforward
\cite{angle}.  The resulting operator is very similar to the operator
used by Penrose in the diagrammatic form of the Spin Geometry Theorem. 
It assigns an angle to two bundles of edges incident to a vertex as
shown in Fig.  \ref{cones}.

The quantum operator is defined as follows:  Incident edges are
partitioned into three categories which correspond to the two cones
shown in Fig.  (\ref{cones}) and the remainder.  All of the edges in
each category connect to a single internal edge in the ``intertwiner
core''.\footnote{When the edges are partitioned into three categories,
as is often convenient in quantum geometry, the external edges are
connected in a branched structure which ends in one principle, internal edge.  The
core of the intertwiner is the trivalent vertex which connects these
three internal edges.  It is the only part of the intertwiner which
must be specified before completing the diagrammatic calculation of
the spectrum.} These edges are denoted $n_{1}$, $n_{2}$, and $n_{r}$. 
For reasons that will be clear in a moment, the spin $n_{r}$ is known
as the ``geometric support''.  Associated to these partitions and
edges are three spin operators $\hat{J}_{1}$, $\hat{J}_{2}$, and
$\hat{J}_{r}$.  Thus,
\begin{equation}
    \begin{split}
	\label{aop}
\hat{\theta}^{(12)}_{v} \ket{s} &:= \arccos \frac{\hat{J}_{1} \cdot \hat{J}_{2}}
{|\hat{J}_{1}| \, | \hat{J}_{2} |} \ket{s} \\
&= {\rm arccos} \left( \frac{n_{r}(n_{r}+2) -
n_{1}(n_{1} +2) - n_{2}(n_{2}+2)} {2 \left[ n_{1}(n_{1} +2 ) \, 
n_{2}(n_{2} +2 ) \right]^{1/2}} \right) \ket{s}.
     \end{split}
\end{equation}
In the second line, the intertwiner core with a trivalent vertex
labeled by $n_{1}$, $n_{2}$ and $n_{r}$ is used.  The key idea of the
angle operator is to measure the relative spins of internal edges 
coming from two disjoint conical regions.

The property which sets this angle operator apart from the other 
geometric observables is the complete absence of the scale of the 
theory $\ell$.  The result is {\it purely} combinatorial!  
However, there is an important technical caveat: If one retains the
diffeomorphism invariance of the classical theory in the construction
of the state space, then there is, generically, a set of continuous
parameters or moduli space associated with higher valence vertices
\cite{moduli}.  These parameters contain information on the embedding
of the spin network graph in space.  On such a space the angle
spectrum is highly degenerate.  It is by no means clear whether this
embedding information is physically relevant \cite{grottthesis}. 
Indeed, the state space of quantum gravity may be described by
abstract or non-embedded graphs.

The spectrum of the angle operator is relatively easy to calculate for
small spins \cite{mikethesis}.  A useful parameter to characterize
this spectrum is the ``spin sum'' $n$, $n = n_{1}+n_{2}+n_{r}$.  The
first few cases are given in Table 1.  Two aspects of the low-lying
spectrum are immediately obvious.  While the examples in Table 1
include 180 degrees, there is a large gap between the smallest angle
and 0 degrees; it is ``hard'' to model small angles.  Second, the
difference between neighboring angles is non-uniform.  This is
especially apparent in the $n=14$ case.  The difference between 0 and
the lowest angle is 50 degrees while neighboring angles are as close
as $ \sim 0.2$ degrees.  Further, neighboring angles do not correspond
to a simple numerical progression of the internal spins $n_{1},
n_{2},$ and $n_{r}$.

The combinatorial nature and the non-uniform spacing of the spectrum
means that the semi-classical limit of the operator offers a wealth of
information about the nature of the spin network vertex.  The
semi-classical limit is not merely a matter of reducing the ``grain''
of the fundamental geometry.  Not only must we match the current
experimental bounds on the smoothness of angle but we must also find
small angles and match the classical distribution of angles in
3-space.

Small angles are `hard' to achieve in the following sense:  To support
an angle near zero the argument of the arc cosine in Eq.  (\ref{aop})
must be close to unity.  That is, we must maximize the argument.  But
the edge labels satisfy the triangle inequalities.  As shown in
Ref.  \cite{mikethesis} a short argument shows that, for a small angle
$\alpha$,
\[
\alpha \geq \frac{4}{\sqrt{n + 8}}.
\]
Since the total spin is bounded from above by twice the spin fluxes
through the cones, the minimum observable angle depends on the area
flux through the regions we are measuring.  For instance, to achieve
an angle as small as $10^{-10}$ radians, the required flux is roughly
$10^{20}$.

\twocolumn[Table 1: Small spin spectra of the angle operator (From 
Ref. \cite{mikethesis}) \\ 
\vspace{0.5 cm}]
\begin{tabular}{c c c | r}
\multicolumn{4}{l}{\textbf{Spin Sum}${} = 4$} \\
$n_{r}$ & $n_{1}$ & $n_{2}$ & $\theta$ (degrees) \\
\hline
2 & 1 & 1 & 70.53 \\
1 & 1 & 2 & 144.74 \\
0 & 2 & 2 & 180.00 \\
\hline
\end{tabular}

Number of distinct angles: 3
 
\vspace{1.5 mm}

\begin{tabular}{c c c | r}
\multicolumn{4}{l}{\textbf{Spin Sum}${} = 6$} \\
$n_{r}$ & $n_{1}$ & $n_{2}$ & $\theta$ (degrees) \\
\hline
3 & 1 & 2 & 65.91 \\
2 & 2 & 2 & 120.00 \\
2 & 1 & 3 & 138.19 \\
1 & 2 & 3 & 155.91 \\
0 & 3 & 3 & 180.00 \\
\hline
\end{tabular}

Number of distinct angles: 5
 
\vspace{1.5 mm}

\begin{tabular}{c c c | r}
\multicolumn{4}{l}{\textbf{Spin Sum}${} = 8$} \\
$n_{r}$ & $n_{1}$ & $n_{2}$ & $\theta$ (degrees) \\
\hline
4 & 2 & 2 & 60.00 \\
4 & 1 & 3 & 63.43 \\
3 & 2 & 3 & 111.42 \\
3 & 1 & 4 & 135.00 \\
2 & 3 & 3 & 137.17 \\
2 & 2 & 4 & 150.00 \\
1 & 3 & 4 & 161.57 \\
0 & 4 & 4 & 180.00 \\
\hline
\end{tabular}

Number of distinct angles: 8
 
\vspace{1.5 mm}

\begin{tabular}{c c c | r}
\multicolumn{4}{l}{\textbf{Spin Sum}${} = 10$} \\
$n_{r}$ & $n_{1}$ & $n_{2}$ & $\theta$ (degrees) \\
\hline
5 & 2 & 3 & 56.79 \\
5 & 1 & 4 & 61.87 \\
4 & 3 & 3 & 101.54 \\
4 & 2 & 4 & 106.78 \\
3 & 3 & 4 & 129.23 \\
4 & 1 & 5 & 133.09 \\
2 & 4 & 4 & 146.44 \\
3 & 2 & 5 & 146.79 \\
2 & 3 & 5 & 156.42 \\
1 & 4 & 5 & 165.04 \\
0 & 5 & 5 & 180.00 \\
\hline
\end{tabular}

Number of distinct angles: 11


\begin{tabular}{c c c | r}
\multicolumn{4}{l}{\textbf{Spin Sum}${} = 12$} \\
$n_{r}$ & $n_{1}$ & $n_{2}$ & $\theta$ (degrees) \\
\hline
6 & 3 & 3 & 53.13 \\
6 & 2 & 4 & 54.74 \\
6 & 1 & 5 & 60.79 \\
5 & 3 & 4 & 96.05 \\
5 & 2 & 5 & 103.83 \\
4 & 4 & 4 & 120.00 \\
4 & 3 & 5 & 124.57 \\
5 & 1 & 6 & 131.81 \\
3 & 4 & 5 & 139.38 \\
4 & 2 & 6 & 144.74 \\
2 & 5 & 5 & 152.34 \\
3 & 3 & 6 & 153.43 \\
2 & 4 & 6 & 160.53 \\
1 & 5 & 6 & 167.40 \\
0 & 6 & 6 & 180.00 \\
\hline
\end{tabular}

Number of distinct angles: 15

\vspace{1.5 mm}

\begin{tabular}{c c c | r}
\multicolumn{4}{l}{\textbf{Spin Sum}${} = 14$} \\
$n_{r}$ & $n_{1}$ & $n_{2}$ & $\theta$ (degrees) \\
\hline
7 & 3 & 4 & 50.77 \\
7 & 2 & 5 & 53.30 \\
7 & 1 & 6 & 60.00 \\
6 & 4 & 4 & 90.00 \\
6 & 3 & 5 & 92.50 \\
6 & 2 & 6 & 101.78 \\
5 & 4 & 5 & 114.46 \\
5 & 3 & 6 & 121.45 \\
6 & 1 & 7 & 130.89 \\
4 & 5 & 5 & 131.08 \\
4 & 4 & 6 & 135.00 \\
5 & 2 & 7 & 143.30 \\
3 & 5 & 6 & 146.05 \\
4 & 3 & 7 & 151.44 \\
2 & 6 & 6 & 156.44 \\
3 & 4 & 7 & 157.79 \\
2 & 5 & 7 & 163.40 \\
1 & 6 & 7 & 169.11 \\
0 & 7 & 7 & 180.00 \\
\hline
\end{tabular}

Number of distinct angles: 19
\onecolumn

\begin{figure}
    \begin{center}
	\pic{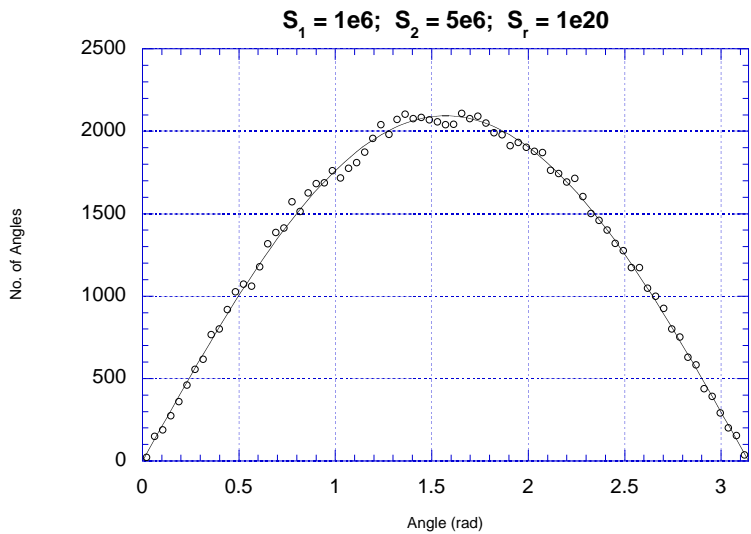}{200}{-10}{-1}{2}
    \end{center}
		 \caption{ \label{dist} The distribution of angles
		 versus angle: The dots represent the distribution of
		 spectral points of the angle operator within a small
		 angle $\delta \theta$.  The solid curve is a plot of
		 $\sin \theta$.  The required spins are $10^{20}$ for
		 geometric support and $10^{6}$ for the surface
		 fluxes. (From Ref. \cite{mikethesis})}
\end{figure}
The angle spectrum must also match the behavior of angles in
three-dimensional space.  In the classical, smooth model of angles the
distribution of solid angles is proportional to $\sin \theta$.  We
expect that the discrete model of space should be able to reproduce
this behavior in the appropriate limit.  Therefore, the classical
limit not only encompasses the need to achieve small angles and fine
angular resolution, but also the need to match the distribution of
solid angles \cite{mikethesis}.  In the discrete model this reduces to
finding the number of intertwiners with a given core.  This
distribution can be found numerically.  The results are shown in Fig. 
(\ref{dist}) for a choice of surface fluxes.  Details of this
calculation are in Ref.  \cite{mikethesis}.

Though it is not obvious by looking at this one plot, to match 
the expected distribution of solid angle it is necessary that the 
geometric support $n_{r}$ be many orders of magnitude larger than the 
surface fluxes through $S_{1}$ and $S_{2}$.  Further, the spin flux 
through the two surfaces must be approximately the same magnitude.

\section{Quasilocal Energy}
\label{quasi}

On spatial manifolds without boundary, the action for the classical
(3+1) form of gravity is a set of constraints.  Suppose that the 
space does have a boundary 
$\partial \Sigma$.  Under the boundary conditions
\begin{equation}
	\label{bcs}
	\begin{split}
		&\delta E^{ai}|_{\partial \Sigma} = 0 \text{ which
		includes fixing the ``area density'' } n_{a} \delta
		E^{ai}|_{\partial \Sigma} = 0 \\
		&N^{a}|_{\partial \Sigma} =0; \; \delta
		N|_{\partial \Sigma} =0
	\end{split}
\end{equation}
the theory is inconsistent without the boundary term
\begin{equation}
	\label{hclass}
H_{\partial \Sigma}(N) = \int_{\partial \Sigma} d^{2}x \epsilon^{ijk} N n_{a}
A_{b}^{i}E^{bj}E^{ak}.
\end{equation}
(See Ref.  \cite{HM} or, for a more general setting, Ref. 
\cite{dis}.)  This is the Hamiltonian for the system.  The
quantization of this observable is carried out in Ref.  \cite{energy}. 
The most general form of the operator is
\begin{equation}
	\label{qeadef}
	\hat{H}_{\partial \Sigma}(N) = -  (4 \pi G) 
	\sum_{v \in S \cap \mathsf{G}} \,  
	\sum_{I,J \dashv v} N_{v}
	\chi_{I} \frac{\hat{J}_{J} \cdot \hat{J}_{I}}
	{\left(\sqrt{ \hat{A}_{v}}\right)^{3}}.
\end{equation}
The first sum is over all the intersections $v$ of the surface with 
the spin network.  The second sum is over all the incident edges $I$ 
and $J$.  $N_{v}$ is the lapse at the intersection $v$; $\chi_{I}$ is 
a sign factor for the edges $I$; and $\hat{A}$ is the area operator.

The spectrum is most easily stated for two types of intersections. 
Type (i) intersections are transversal while type (ii) include all
other intersections.  For higher valence vertices, the intertwiner
core is labeled with edges $p, n$ and $z$ corresponding to incident
edges $I$ with positive, negative and zero values of $\chi_{I}$.  For
the operator defined in Eq.  (\ref{qeadef}), one has for type (i)
\begin{equation}
	\begin {split}
	\label{qeaspec}
	\hat{H}_{\partial \Sigma}(N) \ket{s} 
	&= m \sum_{v \in S \cap {\mathsf G}} 
	N_{v} \sqrt[4]{n_{v}(n_{v}+2)} \ket{s}. \\
\text{For type (ii):}\\
	\hat{H}_{\partial \Sigma}(N) \ket{s}
	&=  m \sum_{v \in S \cap {\mathsf G}}
	N_{v}
	\frac{p_{v}(p_{v}+2) + n_{v}(n_{v}+2) - z_{v}(z_{v}+2)}
	{\left[ 2p_{v}(p_{v}+2) + 2n_{v}(n_{v}+2) - z_{v}(z_{v}+2) 
	\right]^{\frac{3}{4}}} \ket{s}. 
	\end{split}
\end{equation}
The fundamental mass scale is defined as
\[
m := \sqrt{ \frac{\hbar}{ 4 \pi G}}.
\]
This is essentially the same scale as $\ell$ of the geometric observables.

There are two remarks to make: For type (i)
intersections $v$, the quasilocal energy operator has eigenvalue
$\epsilon_{v}$, which is related to the area as $\epsilon_{v} = (m/
\ell) \sqrt{a_{v}}$ in which $a_{v}$ is the eigenvalue of the area
operator.  The quasilocal energy is a sum of contributions from each
intersection; the energy is a sum of independent, non-interacting
particle-like excitations.  This vastly simplifies the statistical
mechanics based on this Hamiltonian \cite{sm}.

\section{Summary}
\label{sum}

There are two new operators for spin net gravity, angle and quasilocal
energy.  With length, area, and volume they provide a number of ways
in which we are able to explore this discrete model of space.  The
angle operator is the first geometric operator that is purely
combinatorial; it depends only on the spins of the vertex, not on the
scale $\ell$.  This also means that it is free of the quantization
ambiguity associated with the Immirzi parameter \cite{I}.  Further,
the quantization of this operator suggests that scalar products also
have discrete spectra.  Though this has largely escaped notice, this
obviously has far reaching consequences in a variety of high and low
energy physics.

In the case of the angle, we learn that the semi-classical limit is
not merely determined by large spins, but by large spins on the
internal edges of the intertwiner located at the vertex.  To correctly
model the distribution of angles we require not only that these spins
be large ($ \sim 10^{6}$) but also that the geometric support be many
orders of magnitude larger ($\sim 10^{20}$).

Both operators share what appears to be a common feature with all
operators in spin net gravity, the spectra are discrete.  What is
utterly remarkable is not just this - the discrete spectra of
geometric quantities - but what this implies for {\it all} of space,
whether it be curved or flat.  If this model is correct, fundamental
quantum geometry affects flat-space physics.  We have only just begun
to realize the consequences of such a dramatic shift in the
foundations of spacetime geometry.

\end{document}